\documentclass[twocolumn,english,showpacs,preprintnumbers,prb]{revtex4-1}
\usepackage{ae,aecompl}
\usepackage[latin9]{inputenc}
\usepackage{color}
\usepackage{amsmath}
\usepackage{amssymb}
\usepackage{graphicx}
\usepackage{esint}

\makeatletter

\usepackage{babel}

\makeatother

\usepackage{babel}
\begin{document}

\title{Spin-dependent beating patterns in thermoelectric
properties: filtering the carriers of the heat flux in a Kondo adatom
system}

\author{A. C. Seridonio$^{1,2}$, E. C. Siqueira$^{2}$, R. Franco$^{3}$,
J. Silva-Valencia$^{3}$, I. A. Shelykh$^{4,5}$, and M. S. Figueira$^{6}$}

\affiliation{$^{1}$Instituto de Geociências e Ciências Exatas - IGCE, Universidade
Estadual Paulista, Departamento de F\'{i}sica, 13506-970, Rio Claro,
SP, Brazil\\
 $^{2}$Departamento de F\'{i}sica e Qu\'{i}mica, Universidade Estadual
Paulista, 15385-000, Ilha Solteira, SP, Brazil\\
 $^{3}$Departamento de Física, Universidad Nacional de Colombia,
A. A. 5997, Bogotá, Colombia\\
 $^{4}$Division of Physics and Applied Physics, Nanyang Technological
University 637371, Singapore\\
 $^{5}$Science Institute, University of Iceland, Dunhagi-3, IS-107,
Reykjavik, Iceland\\
 $^{6}$Instituto de F\'{i}sica, Universidade Federal Fluminense,
24210-340, Niterói, RJ, Brazil}

\begin{abstract}
We theoretically investigate the thermoelectric properties of a
spin-polarized two-dimensional electron gas hosting a Kondo adatom
hybridized with an STM tip. Such a setup is treated within the single-impurity
Anderson model in combination with the atomic approach for the Green's
functions. Due to the spin dependence of the Fermi wavenumbers
the electrical and thermal conductances, together with thermopower and Lorenz number reveal beating patterns as function of the STM tip position in the Kondo regime. In particular, by tuning the lateral displacement of the tip with respect to the adatom vicinity, the temperature and the position of the adatom level, one can change the sign of the Seebeck coefficient through charge and spin. This opens a possibility of the microscopic control of the heat flux analogously to that established for the electrical current.

\end{abstract}


\maketitle

\section{Introduction}

\label{sec1}

In the last few years, the fascinating field
of thermoelectric properties of nano-scale materials is attracting
the growing attention from both experimental \cite{Scheibner_05,Scheibner_07,Scheibner_08,Hoffmann09,Chang,Saira_07,Reddy07}
and theoretical \cite{Dong_02,Nuestro1,Yoshida,Costi} communities
of researchers. In this context, Y. Dubi and M. Di Ventra \cite{DiVentra}
have recently proposed a fundamental setup: two leads connected by a nanoscopic region, in which
thermodynamic quantities such as temperature can be tuned. The possible examples are
quantum dot embedded inside a ballistic channel or a molecule efficiently coupled to both substrate and STM tip
\cite{Reddy07,Baheti08}. The main goal is to achieve a microscopic control of the heat flux analogously
to that performed for the electrical current. Such a task was accomplished in the hybrid
$\text{S-I-N-I-S}$ materials \cite{Saira_07}, where $\text{S}$ stands for the superconducting
leads, $\text{I}$ for the insulating barriers and $\text{N}$ for normal metal. In this device, the heat is carried by the hottest electrons that flow towards the superconductors causing the cooling of the metallic
region. The heat flux is controlled by the voltage applied to an extra lead and it can be increased, decreased or kept constant just by changing this voltage similarly to what is done with electrical current
through an ordinary transistor.

Additionally, novel effects are manifested in the presence
of ferromagnetic leads \cite{SPSTM1,SPSTM2,Kawahara,SPSTM5,Yunong,FM88,FM9,FM10,FM11,FM12,new1,new2,new3} and long spin-relaxation time \cite{SA1,SA2} when thermoelectric properties become spin-dependent. In this case the spin degeneracy of the chemical potentials is lifted and the phenomenon known as spin accumulation arises, thus
affecting the behavior of the thermoelectric quantities.

Another setup promising for the control of thermoelectric flux consists of scanning tunneling microscopy break junction (STMBJ) \cite{Reddy07,Reddy12}.
In this geometry, molecules are trapped between an STM tip of Au kept
at the room temperature $T$ and a substrate of the same material
having different temperature $T+\Delta T$. Molecular junctions are created by moving
the STM tip towards the surface of the metallic electrode and when the circuit is closed, a
bias-voltage is applied and the current is measured.

In the aforementioned systems, the dynamics is ruled by the laws of quantum mechanics, thus leading to wave phenomena analogously to those observed in classical mechanics. In particular, we highlight the so-called beating
effect, which is due to the interference between two waves that propagate in the same
direction with equal amplitudes and slightly different frequencies and wavenumbers. Beating effects appear under certain circumstances in condensed matter physics. As an example they can be detected by using of
the technique of Faraday rotation in CdSe quantum dots: the Zeeman
splitting produces a beating pattern in the spin magnetization \cite{Gupta}. Similar feature is also present in a device composed by two quantum dots coupled to source and drain leads \cite{Trocha2010}.
In such a system, oscillatory gate voltages characterized by slightly different frequencies are attached
to these dots and produce beating pattern in the current signal. Additionally, the appearance of the beats in STM setups has been recently detected in the NbS$_{3}$ one-dimensional
conductor \cite{NbS31,NbS32}.

In this work we focus on the theoretical
study of the thermoelectric properties of spin-polarized two-dimensional
electron gas (2DEG) hosting a Kondo adatom coupled to an STM tip as sketched
in Fig. \ref{fig:Pic1}. The setup is treated by using the
single-impurity Anderson Hamiltonian \cite{SIAM} and the atomic approach
\cite{AP1,Ufinito} for the Green's functions, in which the STM tip and the
``host+adatom'' systems play, respectively, the roles of the cold
and hot reservoirs. In the framework of the linear response theory,
when voltage and temperature gradients are small, we derive analytical
expressions for the thermoelectric coefficients characterizing the system. We find that in the Kondo regime these quantities as functions of the STM tip position exhibit beating patterns, which are due to the dependence of Fermi wavenumbers of the host on spin. We show that in the regime of large Fano factor \cite{Fano1,Fano2}, the thermopower (Seebeck coefficient) alternates
its sign by changing the following degrees of freedom: the lateral separation of the STM tip with respect to the adatom, the temperature and the position of the adatom level. It is worth mentioning that to tune the adatom level with respect to the host Fermi energy, we consider in the model an AFM tip capacitively coupled to this adatom, thus allowing one to control the position of its energy level as originally proposed by some of us in Ref. {[}\onlinecite{Seridonio1}{]}. The cases of presence and absence of the spin accumulation phenomenon are considered. In both of them, positive and negative signs imply that the carriers
responsible for heat conductance are electrons and holes, respectively. Thus we show in this work that the system outlined in Fig. \ref{fig:Pic1} operates as a filter of the spin dependent carriers responsible for the heat conductance.

To our best knowledge, experimental data are not available for the device we consider, but the standard procedure used in the STMBJ experiments should allow experimental verification of our predictions.
It is worth mentioning that STMBJ device usually operates under room temperature, which could be an obstacle for the implementation of such a technique in Kondo regime required for the emergence of the beats. On the other hand, the magnitudes of $T_{K}$ for adatoms are higher with respect to those found in quantum
dots and lie within the range $50\text{{K}}\lesssim T_{K}\lesssim100\text{{K}}$ \cite{Wahl2004}, and thus the observation of the beating patterns in thermoelectric coefficients should not be very complicated experimentally.

It is worth mentioning that the recent experimental findings of Ref.\,{[}\onlinecite{Becker}{]} point out that the STM conductance measurements at $5$K for the 2DEG made by the adsorption of $\text{Cs}$ on the p-doped $\text{InSb(110)}$ surface, in particular under the presences of strong magnetic and electric fields yield an enhanced Rashba effect and consequently, the spin splitting phenomenon that reveals beating patterns in the local density of states (LDOS). These beats are due to the slightly different Fermi wave numbers that appear in the system similarly to ours, thus such results attest the experimental feasibility concerning the catching of beats in STM systems and turn the proposal of this manuscript promising in the same sense.

This paper is organized as follows: in Sec. \ref{sec2}, we develop the theoretical model for the system sketched
in Fig. \ref{fig:Pic1} as well as the derivation of the expressions for thermoelectric coefficients and the Green's function of the Kondo adatom. The results are present in Sec. \ref{sec3} and in Sec.
\ref{sec4}, we summarize our concluding remarks.

\begin{center}
\begin{figure}[!]
\includegraphics[clip,width=0.45\textwidth]{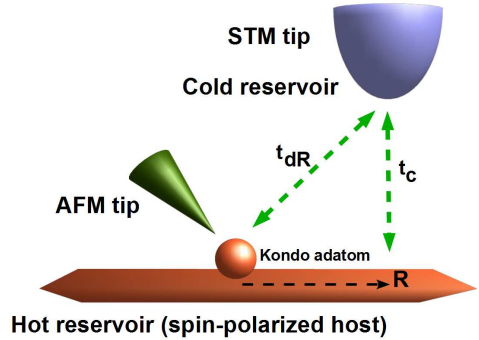}
\caption{\label{fig:Pic1} STM device composed by a normal tip (cold reservoir)
and a Kondo adatom hybridized with a spin-polarized
two-dimensional electron gas (hot reservoir). The parameters $t_{\text{dR}}$
and $t_{\text{c}}$ correspond to the hopping terms in the Hamiltonian. In Kondo regime, the thermoelectric properties characterized by the electrical and thermal conductances, the thermopower (Seebeck coefficient) and Lorenz number entering into the Wiedemann-Franz law exhibit beating patterns if STM tip is displaced laterally with respect to the adatom position. An AFM tip capacitively coupled to the adatom is required as previously proposed in Ref. {[}\onlinecite{Seridonio1}{]} to tune its energy level.}
\end{figure}

\par\end{center}

\section{Theoretical Model}

\label{sec2}

The system we investigate (see Fig.\ref{fig:Pic1}) is described by the following Hamiltonian

\begin{equation}
\mathcal{H}_{\text{{total}}}=\mathcal{H}_{\text{host}}+\mathcal{H}_{\text{tip}}+\mathcal{H}_{\text{tun}},\label{eq:Total}
\end{equation}
where $\mathcal{H}_{\text{host}}$ corresponds to the host electrons in 2DEG and adatom, $\mathcal{H}_{\text{tip}}$ to the STM tip and $\mathcal{H}_{\text{tun}}$ to the tip-host hybridization. In frameworks of the single-impurity Anderson model \cite{SIAM} the terms in Eq. (\ref{eq:Total}) read:
\begin{align}
\mathcal{H}_{\text{host}} & =\sum_{\vec{k}\sigma}\varepsilon_{k\sigma}c_{\vec{k}\sigma}^{\dagger}c_{\vec{k}\sigma}+E_{d}\sum_{\sigma}d_{\sigma}^{\dagger}d_{\sigma}\nonumber \\
 & +V\sum_{\vec{k}\sigma}(c_{\vec{k}\sigma}^{\dagger}d_{\sigma}+\text{{H.c.}})+Ud_{\uparrow}^{\dagger}d_{\uparrow}d_{\downarrow}^{\dagger}d_{\downarrow}.\label{eq:TIAM}
\end{align}

Here the electrons in the host are described by the operator $c_{\vec{k}\sigma}^{\dagger}$
($c_{\vec{k}\sigma}$) for the creation (annihilation) of an electron
in a quantum state labeled by the wave number $\vec{k}$ with an energy

\begin{equation}
\varepsilon_{k\sigma}=\frac{\hbar^{2}k^{2}}{2m}-D_{\sigma},\label{eq:Ek}
\end{equation}
where $D_{\sigma}=D(1+\sigma P)$ is the band half-width in the presence of spin polarization, $P$ is polarization degree of the host defined as
\begin{equation}
P=\frac{\rho_{\text{{host}}}^{\uparrow}-\rho_{\text{{host}}}^{\downarrow}}{\rho_{\text{{host}}}^{\uparrow}+\rho_{\text{{host}}}^{\downarrow}},\label{eq:(SP)}
\end{equation}
where $\rho_{\text{{host}}}^{\sigma}$ are spin dependent densities of states. For the adatom, $d_{\sigma}^{\dagger}$ ($d_{\sigma}$) creates (annihilates) an electron in the state $E_{d}$. Parameter
$V$ describes the hybridization of the adatom with 2DEG. The last term in Eq. (\ref{eq:TIAM}) accounts for the on-site Coulomb interaction $U$.

The Hamiltonian of the tip corresponds to the free electrons described by fermionic operators $b_{\vec{q}\sigma}^{\dagger}$ and $b_{\vec{q}\sigma}$ and reads:
\begin{equation}
\mathcal{H}_{\text{tip}}=\sum_{\vec{q}\sigma}\varepsilon_{q}b_{\vec{q}\sigma}^{\dagger}b_{\vec{q}\sigma}.\label{eq:STM}
\end{equation}

The tunneling Hamiltonian can be expressed as:
\begin{eqnarray}
\mathcal{H}_{\text{tun}} & = & t_{c}\sum_{\vec{q}\sigma}b_{\vec{q}\sigma}^{\dagger}\psi_{R}^{\sigma}+\text{{H.c.}},\label{eq:Tun}
\end{eqnarray}
where $t_{c}$ is the STM tip-host coupling,
\begin{equation}
\psi_{R}^{\sigma}=\sum_{\vec{k}}\phi_{\vec{k}}(\vec{R})c_{\vec{k}\sigma}+(\pi\Delta\rho_{0})^{1/2}qd_{\sigma}\label{eq:PSI_R-1-1}
\end{equation}
is the field operator that accounts for the Fano interference of the tip to 2DEG and tip to adatom paths, $\phi_{\vec{k}}(\vec{R})=e^{i\vec{k}.\vec{R}}$, $\Delta=\pi V^{2}\rho_{0}$ is the Anderson
parameter and $q$ is the Fano factor of the STM device. The latter can be expressed as:
\begin{equation}
q=\frac{t_{dR}}{t_{c}}=q_{0}e^{-k_{F}R},\label{eq:Fano_q}
\end{equation}
where  $t_{\text{dR}}$ and $t_{\text{c}}$ are hopping terms as outlined in Fig. \ref{fig:Pic1}, $k_{F}$ is the Fermi wave number of the host in the case $P=0$ and $R$ is a lateral distance between the tip and the host. Note that according to the Eqs. (\ref{eq:PSI_R-1-1})
and (\ref{eq:Fano_q}), the limit $q_{0}\gg1$ represents the situation in which the tip is highly hybridized with the adatom, while in the opposite regime $q_{0}\ll1$, the tip is strongly connected to the surface {[}see Fig. \ref{fig:Pic1}{]}. Naturally, the increase of the distance between the tip and adatom leads to the quenching of the coupling between them, and for $k_{F}R\gg1$ the Fano parameter drops to zero.

\subsection{Thermoelectric coefficients }

\label{sub:2a}

By applying the linear response theory, and treating tip to host coupling term $\mathcal{H}_{\text{tun}}$ perturbatively, it is possible to show that in absence of spin accumulation  \cite{SA1,SA2}, the charge and spin conductances, $G$ and $G_{S}$, are given by the following expressions:

\begin{equation}
G=G_{\uparrow}+G_{\downarrow}=e^{2}\sum_\sigma I_{o\sigma}\label{G}
\end{equation}
and
\begin{equation}
G_{S}=\frac{e \hbar}{2}\sum_\sigma {\sigma} I_{o\sigma},\label{GS}
\end{equation}where $\sigma=+1$ and $\sigma=-1$ respectively for spin-up and down channels. Similarly, the thermal conductance and  the Seebeck coefficient (thermopower) are given by
\begin{equation}
K=\frac{1}{T}\left(\sum_\sigma I_{2\sigma}-\frac{\left(\sum_\sigma I_{1\sigma}\right)^{2}}{\sum_\sigma I_{o\sigma}}\right)\label{K}
\end{equation}
and
\begin{equation}
S=-\frac{1}{eT}\frac{\sum_\sigma I_{1\sigma}}{\sum_\sigma I_{o\sigma}}, \label{Seebeck}
\end{equation}
where $e>0$ stands for the electron charge. To calculate the transport coefficients $I_{o}$, $I_{1}$, and $I_{2}$,
we follow the paper of B. Dong and X. L. Lei \cite{Dong_02}:

\begin{equation}
I_{n\sigma}=\frac{1}{h}\int\left(-\frac{\partial n_{F}}{\partial\omega}\right)\omega^{n}\tau_{\sigma}(\omega,R)d\omega,\label{L11}
\end{equation}
with

\begin{equation}
\tau_{\sigma}(\omega,R)=\tau_{b}\rho_{\text{LDOS}}^{\sigma}(\omega,R),\label{eq:trans}
\end{equation}
where $h$ is the Planck constant, $n_{\text{{F}}}$ stands for the Fermi-Dirac
distribution, $\tau_{\sigma}(\omega,R)$ is the spin-dependent transmittance,
$\rho_{\text{LDOS}}^{\sigma}(\omega,R)$ is the spin-polarized local
density of states (LDOS) of the ``host+adatom'' system at the position
$\vec{R}$ in the host surface and $\tau_{b}=D_{\sigma}/(1+q^{2})$ is the normalization factor.

For the case of spin accumulation, which is characterized by the lifting of the spin degeneracy in the chemical potentials of the leads, the expressions for the thermal conductance and thermopower
should be modified \cite{SA1,SA2}:

\begin{equation}
\bar{K}=\bar{K}_{\uparrow}+\bar{K}_{\downarrow}=\frac{1}{T}\sum_{\sigma}\left(I_{2\sigma}-\frac{I_{1\sigma}^{2}}{I_{o\sigma}}\right)\label{eq:k2}
\end{equation}
and

\begin{equation}
\bar{S}=\frac{1}{2}({S}_{\uparrow}+{S}_{\downarrow})=-\frac{1}{2eT}\sum_\sigma \frac{I_{1\sigma}}{I_{o\sigma}}.\label{eq:Seebeck2}
\end{equation}

We can also define the spin thermopower $S_{S}$ by the relation

\begin{equation}
S_{S}=\frac{1}{2}({S}_{\uparrow}-{S}_{\downarrow})=-\frac{1}{2eT}\sum_\sigma {\sigma} \frac{I_{1\sigma}}{I_{o\sigma}}.\label{eq:Seebeck22}
\end{equation}

Notice that differently from the case in which there is no spin accumulation
the thermal conductance can be represented as sum of the terms corresponding to spin
up and down channels (compare Eq. (\ref{eq:k2}) and Eq.(\ref{K})).

According to the Wiedemann-Franz (WF) law in ordinary metals, the ratio between the electronic contribution to the thermal conductance $K$ and the temperature $T$ times the electrical conductance $G$ known as Lorenz ratio \cite{Costi},
\begin{equation}
\frac{L}{L_o}\equiv\frac{K(T)}{TG(T)},
\label{LL}
\end{equation}
where $L{o}$ takes an universal value for the Drude gas and is given by  $L_{0}=(\frac{\pi^{2}}{3})(\frac{k_{B}}{e})^{2}$, where $k_{B}$ is the Boltzmann constant and $e$ is the electron charge. In our calculations, the ratio above differs from $L_0$ and together with conductance and Seebeck coefficient reveal beating patterns as function of tip-adatom separation.

In order to obtain the LDOS necessary for the calculation of thermoelectric coefficients, it is convenient to define the retarded Green's function for the field operator in Eq. (\ref{eq:PSI_R-1-1}), which in time domain reads:
\begin{align}
\mathcal{R}_{\psi_{R}\psi_{R}}^{\sigma}\left(t\right) & =-\frac{i}{\hbar}\theta\left(t\right){\tt Tr}\{\varrho[\psi_{R}^{\sigma}\left(t\right),\psi_{R}^{\sigma\dagger}\left(0\right)]_{+}\},\label{eq:PSI_R}
\end{align}
where $\theta\left(t\right)$ is the Heaviside
function, \texttt{Tr} stands for the trace over the Hamiltonian states,  $\varrho$ is the density matrix of the system described by the Hamiltonian {[}Eq. (\ref{eq:TIAM}){]} and $[\cdots,\cdots]_{+}$
stands for the anticommutator. From Eq.~(\ref{eq:PSI_R}), the LDOS of the host can be obtained as
\begin{equation}
\rho_{\text{LDOS}}^{\sigma}(\omega,R)=-\frac{1}{\pi}{\tt Im}(\tilde{\mathcal{R}}_{\psi_{R}\psi_{R}}^{\sigma}),\label{eq:FM_LDOS}
\end{equation}
where $\tilde{\mathcal{R}}_{\psi_{R}\psi_{R}}^{\sigma}$ is the Fourier
transform of $\mathcal{R}_{\psi_{R}\psi_{R}}^{\sigma}(t)$.

To determine an analytical expression for the LDOS, we apply
the equation-of-motion approach. Placing Eq.
(\ref{eq:PSI_R-1-1}) into Eq. (\ref{eq:PSI_R}), one gets:
\begin{align}
\mathcal{R}_{\psi_{R}\psi_{R}}^{\sigma}(t) & =\sum_{\vec{k}\vec{q}}\phi_{\vec{k}}(\vec{R})\phi_{\vec{q}}^{*}(\vec{R})\mathcal{\mathcal{R}}_{c_{\vec{k}}c_{\vec{q}}}^{\sigma}+(\pi\Delta\rho_{0})q^{2}\mathcal{R}_{dd}^{\sigma}\nonumber \\
 & +(\pi\Delta\rho_{0})^{1/2}q\sum_{\vec{k}}[\phi_{\vec{k}}^{*}(\vec{R})\mathcal{R}{}_{dc_{\vec{k}}}^{\sigma}+\phi_{\vec{k}}(\vec{R})\mathcal{R}_{c_{\vec{k}}d}^{\sigma}],\label{eq:GF_1}
\end{align}
which depends on the Green's functions $\mathcal{\mathcal{R}}_{c_{\vec{k}}c_{\vec{q}}}^{\sigma}$,
$\mathcal{R}_{dc_{\vec{k}}}^{\sigma}$, $\mathcal{R}_{c_{\vec{k}}d}^{\sigma}$
and $\mathcal{R}_{dd}^{\sigma}$. First, we have to find
\begin{align}
\mathcal{R}_{c_{\vec{k}}c_{\vec{q}}}^{\sigma}\left(t\right) & =-\frac{i}{\hbar}\theta\left(t\right){\tt Tr}\{\varrho[c_{\vec{k}\sigma}\left(t\right),c_{\vec{q}\sigma}^{\dagger}\left(0\right)]_{+}\}\label{eq:GF_2}
\end{align}
by acting the operator~$\partial_{t}\equiv\partial/\partial t$
on Eq. (\ref{eq:GF_2}). We obtain
\begin{eqnarray}
\partial_{t}\mathcal{R}_{c_{\vec{k}}c_{\vec{q}}}^{\sigma}\left(t\right) & = & -\frac{i}{\hbar}\delta\left(t\right){\tt Tr}\{\varrho[c_{\vec{k}\sigma}\left(t\right),c_{\vec{q}\sigma}^{\dagger}\left(0\right)]_{+}\}\nonumber \\
 & - & \frac{i}{\hbar}\varepsilon_{k}\mathcal{R}_{c_{\vec{k}}c_{\vec{q}}}^{\sigma}\left(t\right)-\frac{i}{\hbar}V\mathcal{R}_{dc_{\vec{q}}}^{\sigma}\left(t\right),
\label{eq:GF_3}
\end{eqnarray}
where we used that
\begin{align}
i\hbar\partial_{t}c_{\vec{k}\sigma}\left(t\right) & =[c_{\vec{k}},\mathcal{H}_{\text{host}}]=\varepsilon_{k}c_{\vec{k}\sigma}+Vd_{\sigma}\left(t\right).\label{eq:HB_I}
\end{align}

In the energy domain $\omega$, we solve Eq. (\ref{eq:GF_3}) for
$\tilde{\mathcal{R}}_{c_{\vec{k}}c_{\vec{q}}}^{\sigma}$ and obtain
\begin{align}
\tilde{\mathcal{R}}_{c_{\vec{k}}c_{\vec{q}}}^{\sigma} & =\frac{\delta_{\vec{k}\vec{q}}}{\varepsilon^{+}-\varepsilon_{k}}+\frac{V}{\varepsilon^{+}-\varepsilon_{k}}\tilde{\mathcal{R}}_{dc_{\vec{q}}}^{\sigma},\label{eq:GF_4}
\end{align}
where $\varepsilon^{+}=\omega+i\eta$ and $\eta\rightarrow0^{+}$.
Notice that we also need to calculate the mixed Green's function $\tilde{\mathcal{R}}_{dc_{\vec{q}}}.$
Analogously, we find
\begin{equation}
\tilde{\mathcal{R}}_{dc_{\vec{q}}}^{\sigma}=\frac{V}{\varepsilon^{+}-\varepsilon_{q}}\tilde{\mathcal{R}}_{dd}^{\sigma}=\tilde{\mathcal{R}}_{c_{\vec{q}}d}^{\sigma}.\label{eq:GF_9}
\end{equation}

Now within the wide band limit $D\rightarrow\infty$, we place Eq.
(\ref{eq:GF_9}) into Eq. (\ref{eq:GF_4}) and then substitute these equations back into Eqs. (\ref{eq:FM_LDOS})
and (\ref{eq:GF_1}). This procedure results into the following expression for the spin-polarized LDOS of the system:
\begin{align}
\rho_{\text{{LDOS}}}^{\sigma}(\omega,R) & =\rho_{\text{{host}}}^{\sigma}+\rho_{0}\Delta[(\mathcal{F}_{\sigma}^{2}-q^{2}){\tt Im}(\tilde{\mathcal{R}}_{dd}^{\sigma})\nonumber \\
 & +2\mathcal{F}_{\sigma}q{\tt Re}(\tilde{\mathcal{R}}_{dd}^{\sigma})]\label{eq:LDOS_p1}
\end{align}
where
\begin{align}
\mathcal{F}_{\sigma} & =\frac{1}{\rho_{0}}\sum_{\vec{k}}\phi_{\vec{k}}(\vec{R})\delta(\varepsilon-\varepsilon_{k\sigma})=\frac{\rho_{\text{{host}}}^{\sigma}}{\rho_{0}}J_{0}(k_{\text{{F}}\sigma}R)\label{eq:Friedel_j}
\end{align}
accounts for the Friedel oscillations described by the \textit{zeroth}
order Bessel function $J_{0}$ dependent on the spin-dependent
Fermi wavenumbers as follows:
\begin{equation}
k_{\text{{F}}\downarrow}=\sqrt{\frac{1-P}{1+P}}k_{\text{{F}}\uparrow} , \label{eq:Fs}
\end{equation}
where in all the figures we choose $k_{F}=k_{F\uparrow}$ and $k_{F\downarrow}$ is calculated employing the above equation.

Additionally, to determine the LDOS, we need to find the Green's function
$\tilde{\mathcal{R}}_{dd}^{\sigma}$ of the adatom. In the present work it is obtained via the atomic approach in the limit of
infinite on-site Coulomb interaction.

\subsection{The atomic approach}

\label{sub:2b}

In order to implement the atomic approach for the case of the infinite Coulomb energy \cite{AP1},
we begin with Eq. (\ref{eq:TIAM}) expressed as
\begin{eqnarray}
\mathcal{H}_{\text{{host}}} & =
\displaystyle\sum_{\vec{k}\sigma}\varepsilon_{k\sigma}c_{\vec{k}\sigma}^{\dagger}c_{\vec{k}\sigma}+E_{d}\displaystyle\sum_{\sigma}X_{d,\sigma\sigma}\nonumber \\
 & +V\displaystyle\sum_{\vec{k}\sigma}(c_{\vec{k}\sigma}^{\dagger}X_{d,0\sigma}+\text{{H.c.}}),\label{SIAM2}
\end{eqnarray}
where $X_{p,ab}=|p,a\rangle\langle p,b|$ is the Hubbard operator that projects out the doubly occupied state from the adatom to ensure the limit of infinite Coulomb correlation, the label $(a,b)$
defines the parameters associated with the corresponding atomic transition. This formalism is based on an extension of the Hubbard cumulant expansion also applicable to the Anderson lattice with impurity-host
couplings treated as perturbations. The use of this expansion allows one to express the
exact Green's function in terms of an unknown effective cumulant. In previous works \cite{AP1,Ufinito}, we have studied the Anderson impurity with an approximate effective cumulant obtained from the atomic limit of the model in a procedure
that we call the zero band width (ZBW) approximation.

As we are interested in the exact Green's function for the adatom, we use the standard definition
\begin{equation}
\mathcal{R}_{dd}^{\sigma}\left(t\right)=-\frac{i}{\hbar}\theta\left(t\right){\tt Tr}\{\varrho[d_{\sigma}\left(t\right),d_{\sigma}^{\dagger}\left(0\right)]_{+}\}.\label{eq:G_d_d}
\end{equation}

The Fourier transformation of Eq. (\ref{eq:G_d_d}) over time coordinate provides
the adatom Green's function in energy domain, which is then obtained by replacing the bare cumulant by the effective
one calculated by following the atomic approach with the Hamiltonian
in Eq. (\ref{SIAM2}). As a result, we have
\begin{equation}
\tilde{\mathcal{R}}_{dd}^{\sigma}(\omega)=\dfrac{\mathcal{M}_{\text{{eff}}}^{\sigma}(\omega)}
{1-\mathcal{M}_{\sigma}^{eff}(\omega)|V|^{2}\sum_{\vec{k}}\mathcal{\tilde{\mathcal{R}}}_{c}^{\sigma}(\vec{k},\omega)},\label{Eq.6}
\end{equation}
for  the adatom Green's function in terms of the effective
cumulant $\mathcal{M}_{\text{{eff}}}^{\sigma}(\omega)$ and the free-electron
Green's function
\begin{equation}
\mathcal{\tilde{\mathcal{R}}}_{c}^{\sigma}(\vec{k},\omega)=\frac{1}{\omega-\varepsilon_{\vec{k}\sigma}+i\eta},\label{eq:AA_freeGF}
\end{equation}
where $\eta\to0^{+}$. The atomic version of Eq. (\ref{Eq.6}) is
given by:
\begin{equation}
\mathcal{\tilde{R}}_{dd,\text{{at}}}^{\sigma}(\omega)=\frac{\mathcal{M}_{\text{{at}}}^{\sigma}(\omega)}{1-\mathcal{M}_{\text{{at}}}^{\sigma}(\omega)|V|^{2}\mathcal{\tilde{R}}_{\text{{ZBW}}}^{\sigma}(\omega)},\label{Eq.6a}
\end{equation}
which results in
\begin{equation}
\mathcal{M}_{\text{{at}}}^{\sigma}(\omega)=\frac{\mathcal{\tilde{R}}_{dd,\text{{at}}}^{\sigma}(\omega)}{1+\mathcal{\tilde{R}}_{dd,\text{{at}}}^{\sigma}(\omega)|V|^{2}\mathcal{\tilde{R}}_{\text{{ZBW}}}^{\sigma}(\omega)},\label{Eq.7}
\end{equation}
for the effective cumulant determined from the adatom Green's
function, both dependent on
\begin{equation}
\mathcal{\tilde{R}}_{\text{{ZBW}}}^{\sigma}(\omega)=\frac{1}{\omega-(\epsilon_{0\sigma}-\mu)+i\eta},\label{Eq3.144}
\end{equation}
for an electron state, in the ZBW approximation with $\mu$ as
the chemical potential of the host. As one can see, Eq. (\ref{Eq3.144})
replaces all energy contributions of the original Fermi sea by two
spin dependent atomic levels, i.e., one can perform the substitution $\sum_{\vec{k}\sigma}\varepsilon_{k\sigma}c_{\vec{k}\sigma}^{\dagger}c_{\vec{k}\sigma}\rightarrow\sum_{\sigma}\epsilon_{0\sigma}c_{0\sigma}^{\dagger}c_{0\sigma}$ in Eq. (\ref{SIAM2}) with $\epsilon_{0 \uparrow} = (1+P)\varepsilon_{0 \uparrow}$ and $\epsilon_{0 \downarrow} = (1-P)\varepsilon_{0 \downarrow}$
representing the band atomic levels  corresponding to each one of the spin polarized conduction bands (for more details see the Appendix of the atomic approach work in Ref. [\onlinecite{AP1}]). The ZBW overestimates the conduction electrons contribution concentrating them at a single energy level $\epsilon_{0\sigma}$, and to moderate this effect we shall replace $V^{2}$ by $\Delta_{\sigma}^{2}$ in Eqs. (\ref{Eq.6a})
and (\ref{Eq.7}), where $\Delta_{\sigma}=\pi V^{2}/2D_{\sigma}$
is the spin-dependent Anderson parameter.

To determine the adatom Green's function, we use
the atomic cumulant $\mathcal{M}_{\text{{at}}}^{\sigma}(\omega)$
in Eq. (\ref{Eq.6}) and verify that
\begin{equation}
\mathcal{\tilde{R}}_{dd}^{\sigma}(\omega)=\frac{\mathcal{M}_{\text{{at}}}^{\sigma}(\omega)}{1-\mathcal{M}_{\text{{at}}}^{\sigma}(\omega)\dfrac{\left\vert V\right\vert ^{2}}{2D_{\sigma}}\ln\left[\dfrac{\omega-D_{\sigma}+\mu}{\omega+D_{\sigma}+\mu}\right]},\label{Gff_ap}
\end{equation}
which provides an analytical expression in the flat band approximation. The Stoner splitting was not considered in this equation, just that arising from the exchange field of the host, since the former breaks the particle-hole symmetry of the conduction band in this host and prevents the employment of the Friedel's sum rule in the current version of the atomic approach. In principle, we could extend the Friedel's sum rule to account the Stoner splitting, but the spin splitting of the Kondo peak arises from the aforementioned exchange field, which is the most relevant effect that defines, in combination with the slightly different spin-dependent Fermi wave numbers of the host, the behavior of the thermoelectric properties of the system.

As the final step we have to find the proper values of the effective atomic levels $\varepsilon_{0\sigma}$
that well describe the ZBW Green's functions in Eq. (\ref{Eq3.144}) and consequently, the adatom Green's
function. To that end, we use the condition that in metallic systems the most important region in the energy range for conduction electrons is located at the chemical potential $\mu$ and that the Friedel's sum rule is satisfied \cite{Friedel} for the adatom spectral density:
\begin{equation}
\rho_{d,\sigma}(\mu)=-\frac{1}{\pi}\text{{Im}}[\tilde{\mathcal{R}}_{dd}^{\sigma}(\mu)]=\frac{\sin^{2}[\delta_{\sigma}(\mu)]}{\pi\Delta_{\sigma}},\label{fried}
\end{equation}
where $\delta_{\sigma}(\mu)=\pi n_{d,\sigma}$
is the conduction phase shift at the chemical potential,  and $n_{d,\sigma}$ is the spin dependent adatom occupation. We can thus calculate self-consistently the atomic levels $\varepsilon_{0\sigma}$  using Eq. (\ref{fried}) together with the relation
\begin{equation}
n_{d,\sigma}=\langle X_{d,\sigma\sigma}\rangle=-\frac{1}{\pi}\int_{-\infty}^{+\infty}n_{F}\text{{Im}}(\mathcal{\tilde{R}}_{dd,\text{{at}}}^{\sigma}(\omega))d\omega.\label{G00}
\end{equation}

\begin{figure}[!]
\includegraphics[width=0.4\textwidth,angle=-90.]{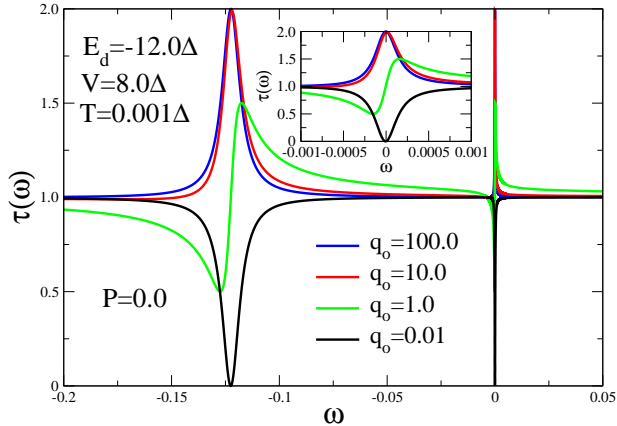} \caption{\label{fig:Pic2}(Color online) The plot of total transmittance $\tau(\omega)=\tau_{\uparrow}(\omega,R=0)+\tau_{\downarrow}(\omega,R=0)$ entering into Eq. (\ref{eq:trans}) as function of $\omega$ for several representative values for $q_{0}$: $q_{0}=100.0$, $q_{0}=10.0$, $q_{0}=1.0$ and $q_{0}=0.01$. In the  Kondo regime, it exhibits two characteristic peaks: the resonance due to the localized level $E_{d}$ in the domain $\omega<0$ and the Kondo peak placed at the chemical potential $\omega=\mu=0$ of the host. The other values for  $q_{o}$ show the crossover from the Kondo limit towards the Fano antiresonance regime, which is observed with $q_{o}=0.01$. The inset shows the behavior of the transmittance in the vicinity of $\mu=0$. Off the resonances, the transmittance decays to unitary value of the background contribution. The values of the parameters are: $P=0$ (non magnetic host), $E_{d}=-12\Delta$, $V=8.0\Delta$ and $T=0.001\Delta$.}
\end{figure}

Rigorously speaking, the Friedel's sum rule is only valid for the temperature $T=0$K, but we employ it as an approximation at temperatures below or in the same order of the Kondo temperature by determining the parameter $\varepsilon_{0\sigma}$ in Eq. (\ref{Eq3.144}) to fix the Kondo peak at the chemical potential $\mu=0$. As in the atomic approach a closed expression for $T_{K}$ is unknown, its definition can be performed qualitatively just by imposing the half-width of the Kondo peak as an approximated measure of such a quantity. For the case of the parameters employed in the current calculation, we obtain from the LDOS of Fig. \ref{fig:Pic2} with $q_{o}=10.0$ that $T_{K} \simeq 0.002\Delta,$ thus ensuring the applicability of the method. We emphasize that the fulfillment of the Friedel's sum rule is indispensable to describe the Kondo peak below $T_{K}$, since at high temperatures as $T \geq \Delta,$ the system is driven to a regime where the Friedel's sum rule breaks and the Kondo peak no longer exists. For this situation, the atomic approach recovers the standard Hartree-Fock description of the single impurity Anderson model \cite{Doniach}.

Additionally, we stress that the Friedel's sum rule in combination with the atomic approach enclose exclusively local calculations as that of the adatom Green's function, which is a spatial independent quantity according to Eq. (\ref{Eq.6a}). As a result, it ensures that the energy excitations in the host conduction band do not depend on this degree, thus preventing an oscillatory behavior by means of the impurity. Indeed, such a signature arises from Friedel oscillations in the 2DEG, being assisted by slightly different spin-dependent Fermi wave numbers, which are the source of the beats in the LDOS as we will discuss in the next section. These features can be confirmed looking at Eqs. (\ref{eq:LDOS_p1}), (\ref{eq:Friedel_j}) and (\ref{eq:Fs}), where we can clear visualize an explicit dependence on the STM-tip position pulled out from the Green's function of the adatom.

By employing the atomic approach, we can find the transmittance $\tau(\omega)=\tau_{\uparrow}(\omega,R=0)+\tau_{\downarrow}(\omega,R=0)$ (see Eq. (\ref{eq:trans})) as a function of the single particle energy $\omega$. For the case of the STM tip placed right above the adatom $(R=0)$ the result is shown at Fig. \ref{fig:Pic2}. The Fano-Kondo behavior is ruled by the parameter $q_{o}$ defined in Eq. (\ref{eq:Fano_q}). For $q_{o} >> 1$ we have the Kondo limit, while $q_{o} << 1$ leads to the Fano antiresonance regime. To perceive this feature, we plot several representative values of the parameter $q_{o}$ in Fig. \ref{fig:Pic2}, just in order to verify the crossover from the Kondo limit $q_{o}=100.0$ towards the Fano antiresonance regime established by $q_{o}=0.01$. In the inset of the same figure, we show in detail such a crossover in the vicinity of $\mu=0$. As we employed in the calculations $q_{o}=10.0$, the transmittance is characterized by the Kondo peak, but with a small fingerprint of the Fano effect. Moreover, off the resonances, the transmittance approaches the unitary value of the background contribution, which arises from the conduction band of the metallic surface. This confirms that the atomic approach is a reliable technique to capture the many-body physics of the Kondo effect, which allows us to safely apply it to the analysis of the thermoelectric properties of the setup presented in the next section.

\begin{figure}[!]
\includegraphics[width=0.4\textwidth,angle=-90.]{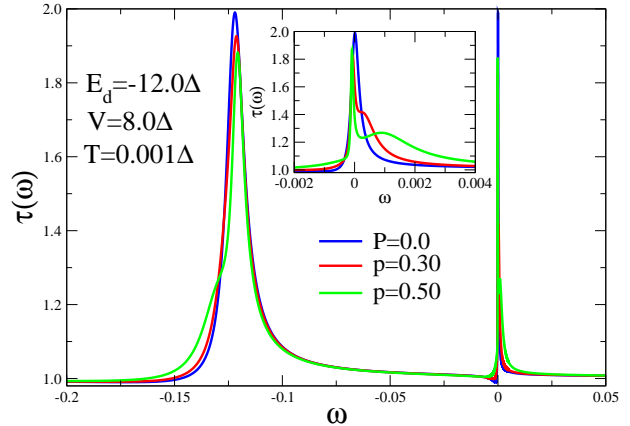} \caption{\label{fig:Pic22} (Color online) Total transmittance $\tau(\omega)=\tau_{\uparrow}(\omega,R=0)+\tau_{\downarrow}(\omega,R=0)$ entering into Eq. (\ref{eq:trans}) as function of $\omega$ for several representative $P$ values: $P=0$, $P=0.30$ and $P=0.50$. The values of the parameters are: $q_{o}=10.0$,  $E_{d}=-12\Delta$, $V=8.0\Delta$ and $T=0.001\Delta$}
\end{figure}

In Fig. \ref{fig:Pic22} we plot the total transmittance $\tau(\omega)$ as function of $\omega$ for several values of $P$: $P=0$, $P=0.30$ and $P=0.50$. In the Kondo regime, we can observe two characteristic peaks: the broader resonance is due to the localized level $E_{d}$ of the adatom, while the sharper is the Kondo peak placed at the chemical potential $\omega=\mu=0$ of the host. As the polarization increases, the spin-up and down channels become resolved, thus yielding two satellite structures around $\mu=0$ as can be clearly visualized in the inset for the case  $P=0.50$.

\section{Results}

\label{sec3}

\begin{figure}[!]
\includegraphics[width=0.4\textwidth,angle=-90.]{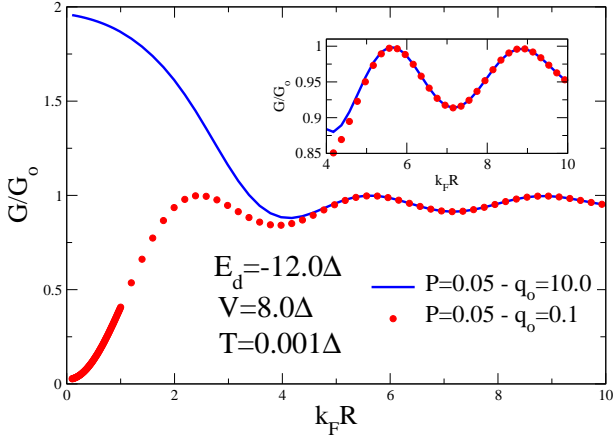} \caption{\label{fig:Pic3}(Color online) (a) Electrical conductance $G$ as function of tip-adatom separation in Fano-Kondo regime. For $q_{0}=10$ constructive Fano interference combined with Kondo effect leads to the appearance of the conductance maximum at $k_{F}R=0$. Contrastingly, in the case of $q_{0}=0.10$ the Fano interference is destructive, which leads to the conductance minimum at $k_{F}R=0$. In both cases, Friedel oscillations are clearly seen for finite values of $k_{F}R$ and reveal universal pattern independent on $q$ as it is shown at the inset. The values of the parameters are $P=0.05$, $E_{d}=-12\Delta$, $V=8.0\Delta$ and $T=0.001\Delta$.}
\end{figure}
In this section we present the results for thermoelectric coefficients characterizing the system keeping the values of the parameters used in Fig. \ref{fig:Pic2} and shown in the corresponding caption.

In Fig. \ref{fig:Pic3} we show the electrical conductance $G$ of Eq. (\ref{G}) in units of $G_{o}=e^{2}/h$ as a function of $k_{F}R$. We compare the behaviors of $G$ for the same value of host polarization $P$ and two different values of Fano parameter $q_{o}=10.0$ and $q_{o}=0.1$. In the former case, the electrical conductance at $k_{F}R$=0 remains close to $G/G_{o}=2$, since the STM device acts as a single electron transistor \cite{QD1,QD2}. On the other hand, for $q_{o}=0.1$ due to the destructive interference the electrical conductance is completely suppressed in analogy to that observed in T-shaped quantum dots \cite{ACS09}. For $k_{F}R>5$, spin-polarized Friedel oscillations manifest, their shape is independent on $q$ and is ruled exclusively by the polarization $P$ as it is shown at the inset of Fig. \ref{fig:Pic3}.

\begin{figure}[!]
\includegraphics[width=0.4\textwidth,angle=-90.]{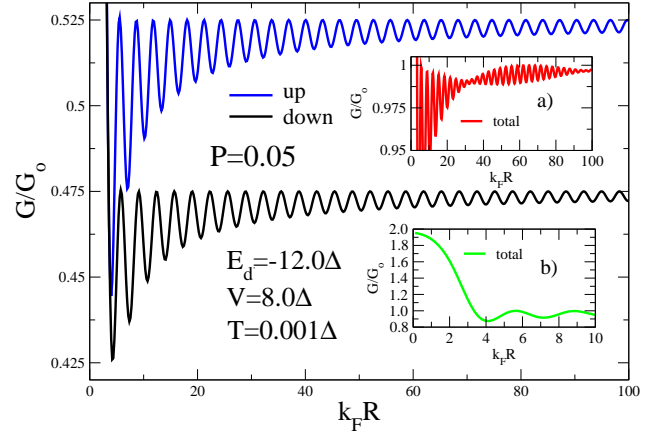} \caption{\label{fig:Pic4}(Color online) Spin-resolved electrical conductances $G_{\uparrow}$ and $G_{\downarrow}$ in the Kondo regime as function of tip-adatom separation. Due to the non-zero polarization of the host ($P=0.05$) the conductance in spin-up channel is bigger than the conductance in spin-down channel. Both channels reveal spin dependent Friedel oscillations. The used parameters are: $q_{0}=10$, $P=0.05$ (ferromagnetic host), $E_{d}=-12\Delta$, $V=8.0\Delta$ and $T=0.001\Delta$. The insets for $G=G_{\uparrow}+G_{\downarrow}$ show the regime of the large distances where beating pattern is clearly observed (inset (a)) and small distances, where this pattern is absent (inset (b)).}
\end{figure}

To better understand the spin-polarized Friedel oscillations, we split the electrical conductance $G$ into spin resolved parts $G_{\uparrow}$ and $G_{\downarrow}$ as it is displayed in Fig. \ref{fig:Pic4} for $P=0.05$. As one can see the spin-up component is shifted towards higher values of $G$ and spin-down component moves in the opposite direction. This is due the spin-dependence of LDOS entering into Eqs. (\ref{G}), (\ref{L11}) and (\ref{eq:trans}). The difference in the Fermi wavenumbers for spin-up and spin-down electrons $k_{F\uparrow}$ and $k_{F\downarrow}$ results in a slight difference of the frequencies of the oscillations for spin resolved components of the conductance, which leads to the onset of the beating pattern in the total conductance shown at the inset (a) in the region of large tip-adatom separations.  In the range of small distances between adatom and STM tip, such a feature does not emerge as it is seen at the inset (b).

\begin{figure}[!]
\includegraphics[width=0.4\textwidth,angle=-90.]{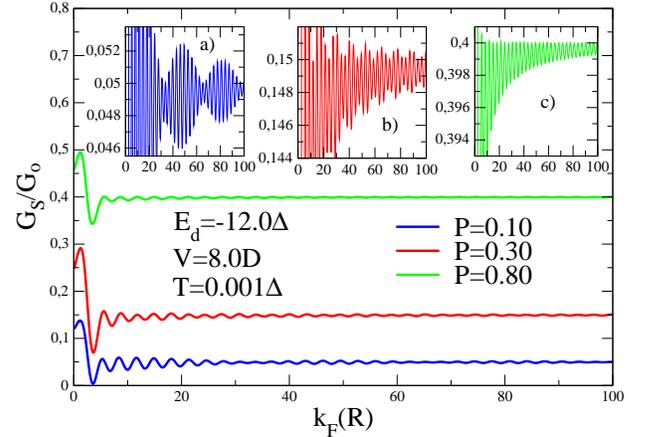} \caption{\label{fig:Pic13}(Color online) Spin electrical conductance $G_{S}$ in the Kondo regime as function of  tip-adatom separation for the polarization $P=0.10$ $(\frac{k_{F\downarrow}}{k_{F\uparrow}}=0.9)$, $P=0.30$ $(\frac{k_{F\downarrow}}{k_{F\uparrow}}=0.73)$ and $P=0.80$ $(\frac{k_{F\downarrow}}{k_{F\uparrow}}=0.33)$. For lower polarizations, the beating pattern is present, but as the  polarization increases such a pattern disappears gradually. The used parameters are: $q_{0}=10.0$, $E_{d}=-12\Delta$, $V=8.0\Delta$ and $T=0.001\Delta$. The inset shows the beats for $P=0.10,$ $P=0.30$ and $P=0.80.$}
\end{figure}

In Fig. \ref{fig:Pic13} we plot the spin electrical conductance $G_{S}$ of Eq. (\ref{GS}) within the Kondo regime as function of the tip-adatom separation for several values of $P$. Near the adatom site, $G_{S}$ presents strong oscillations and faraway from this position, in particular for the situation of lower polarizations as $P=0.10$, a beating pattern is verified in the $G_{S}$ profile. In the insets (a), (b) and (c) we present the evolution of the beats corresponding to the curves of the main plot. For $P=0.10$ (inset (a)), the beating pattern is present, while for $P=0.30$ it no longer exists but the curve of the inset (b) still preserves some structure of the beats. In the case of  $P=0.80,$ such a pattern is completely absent as we can verify in panel (c). Thereby, this behavior with increasing $P$ shows that Eq. (\ref{eq:Fs}) holds and that the slightly different spin-dependent Fermi numbers are the underlying mechanism for the beats formation. By increasing $P$, $G_{S}$ is positive and obeys the same trend of $G_{\uparrow},$ which becomes much higher than $G_{\downarrow},$ otherwise $G_{S}$ would be negative as a result of the inequality $G_{\uparrow}<G_{\downarrow}.$ Consequently, the present results due to Eq. (\ref{GS}) point out that the system can operate as a spin-filter.

\begin{figure}[!]
\includegraphics[width=0.4\textwidth,angle=-90.0]{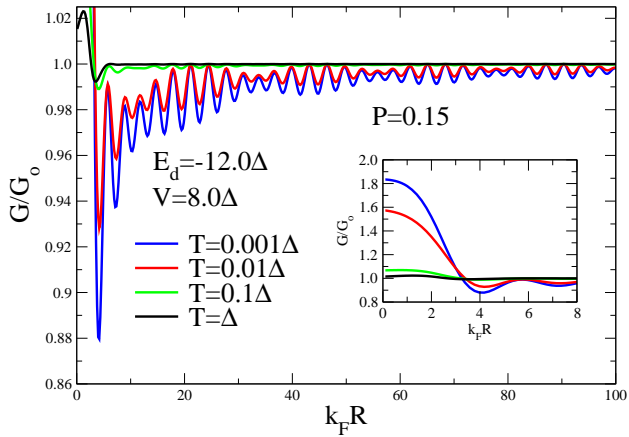} \caption{\label{fig:Pic5}(Color online) Total conductance as function of the adatom-tip separation presented for different values of the temperature. For temperatures above Kondo temperature the beating pattern is suppressed. The inset shows the behavior of the conductance for small tip-adatom separations. The used values of the parameters are: $q_{0}=10$, $P=0.15$ (magnetic host), $E_{d}=-12\Delta$ and $V=8.0\Delta$.}
\end{figure}

\begin{figure}[!]
\includegraphics[width=0.4\textwidth,angle=-90.]{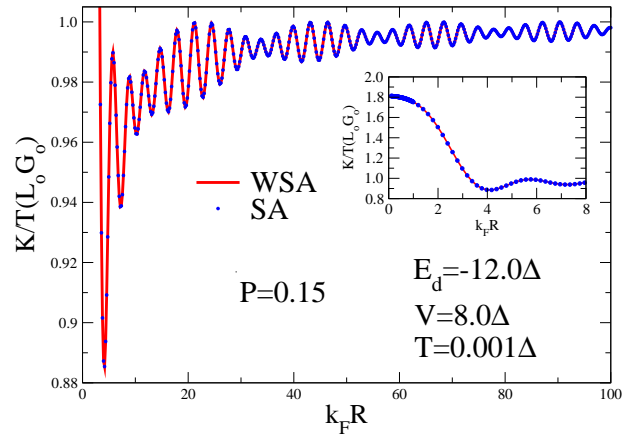} \caption{\label{fig:Pic6}(Color online) Thermal conductance $K$ over temperature in $L_{o}G_{o}$ units, in the Kondo regime as function of tip-adatom separation. Both cases of the absence of spin accumulation (denoted as WSA) and the presence of spin accumulation (denoted as SA) are presented and show very similar behavior. The used parameters are: $q_{0}=10$, $E_{d}=-12\Delta$, $V=8.0\Delta$, $T=0.001\Delta$ and $P=0.15$.}
\end{figure}

\begin{figure}[!]
\includegraphics[width=0.4\textwidth,angle=-90.]{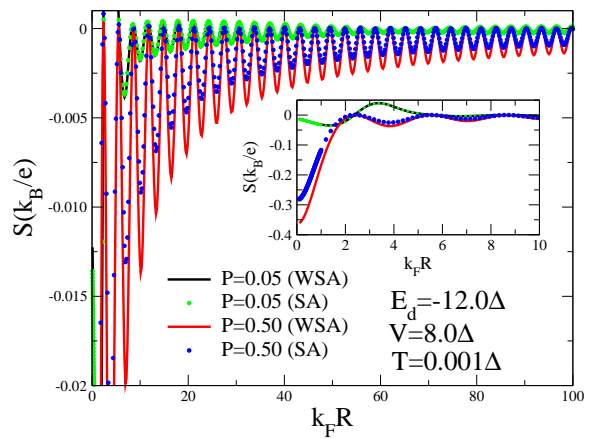} \caption{\label{fig:Pic66}(Color online) Thermopower $S$ in the Kondo regime as function of tip-adatom separation. Both cases of the absence of spin accumulation (denoted as WSA) and the presence of spin accumulation (denoted as SA) are presented. The used parameters are: $q_{0}=10$, $E_{d}=-12\Delta$, $V=8.0\Delta$, $T=0.001\Delta$ and $P=0.05$ and $P=0.50$.}
\end{figure}

In Fig. \ref{fig:Pic5} we present the electrical conductance $G/G_{o}$ of Eq. (\ref{G}) as a function of  $k_{F}R$ with $P=0.15$ and $q_{o}=10$ for different temperatures $T$. The plot reveals that the beating pattern only appears at temperatures below the Kondo temperature $T_K$. As it was discussed in introduction, the characteristic values of $T_K$ lie in the range $50K\lesssim T_{K}\lesssim100K$ and this regime is thus easily accessible experimentally. The inset of Fig. \ref{fig:Pic5}, shows the behavior of the electrical conductance for small tip-adatom separations.  One can clearly see that Kondo effect dominates the tunneling through the adatom leading to the enhancement of the conductance. As temperature increases Kondo effect disappears and the conductance reaches the value given by the background contribution from the host surface.

Fig. \ref{fig:Pic6} shows the thermal conductances over temperature of Eqs. (\ref{K}) and (\ref{eq:k2}) measured in the units of $L_{o}G_{o}.$ We consider both cases of absence and presence of spin-accumulation and show that for $P=0.15$ the results are almost the same in both situations. However, in the situation of a large polarization $P$, the scenarios with and without the spin-accumulation effect, respectively identified by the labels SA and WSA, lead to distinguishable results. As the thermopower is more susceptible to such a phenomenon than other thermoelectric properties, we can see that Fig. \ref{fig:Pic66} reveals two distinct behaviors arising from $P=0.05$ and $P=0.50,$ in which only the latter shows the SA and WSA cases resolved. The present feature thus ensures that the scenarios SA and WSA can deviate from each other just by increasing $P.$

\begin{figure}[!]
\includegraphics[width=0.4\textwidth,angle=-90.]{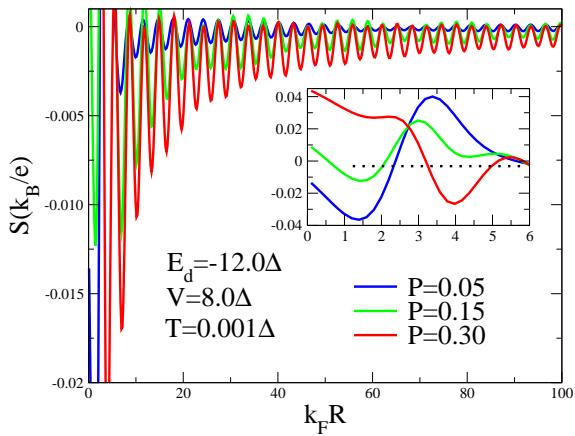} \caption{\label{fig:Pic7}(Color online) Seebeck coefficient (thermopower) $S$  of Eq. (\ref{Seebeck}) in the Kondo regime as function of tip-adatom separation for different values of the polarization of the host. The inset shows the behavior of $S$ for small tip-adatom separations, where $S$ exhibits the alternation of a sign. The heat flux is transmitted mainly by electrons in the region $S<0$ and mainly by holes in the region $S>0$. The used parameters are: $q_{0}=10$, $E_{d}=-12\Delta$, $V=8.0\Delta$, $T=0.001\Delta$ and different values of the spin-polarization $P$.}
\end{figure}

\begin{figure}[!]
\includegraphics[width=0.4\textwidth,angle=-90.]{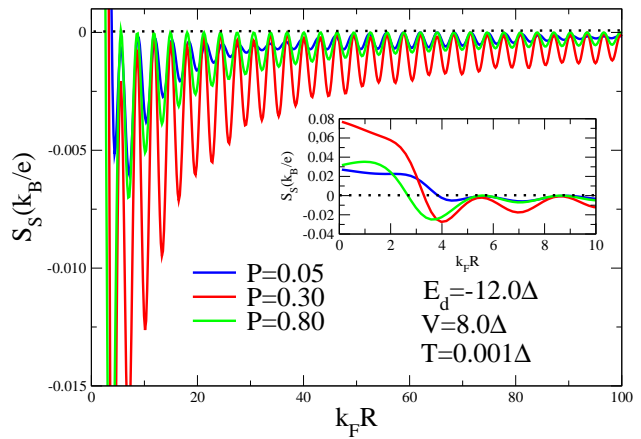} \caption{\label{fig:Pic14}(Color online) Spin Seebeck coefficient (thermopower) $S_{S}$ of Eq. (\ref{eq:Seebeck22}) in the Kondo regime as function of tip-adatom separation for several polarization values. The used parameters are: $q_{0}=10$, $E_{d}=-12\Delta$, $V=8.0\Delta$ and $T=0.001\Delta$. The inset shows the profiles of such quantities near the adatom.}
\end{figure}

\begin{figure}[!]
\includegraphics[width=0.4\textwidth,angle=-90.]{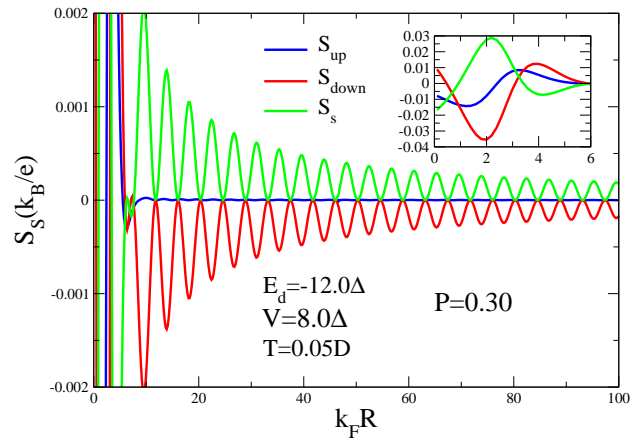} \caption{\label{fig:FFig2}(Color online) Spin Seebeck coefficient (thermopower) $S_{S}$ of Eq. (\ref{eq:Seebeck22}) and spin components $S_{\sigma}$ in the Kondo regime as function of tip-adatom separation for $P=0.3.$ The used parameters are: $q_{0}=10$, $E_{d}=-12\Delta$, $V=8.0\Delta$ and $T=0.05\Delta$. The inset shows the profiles of such quantities near the adatom.}
\end{figure}

\begin{figure}[!]
\includegraphics[width=0.4\textwidth,angle=-90.]{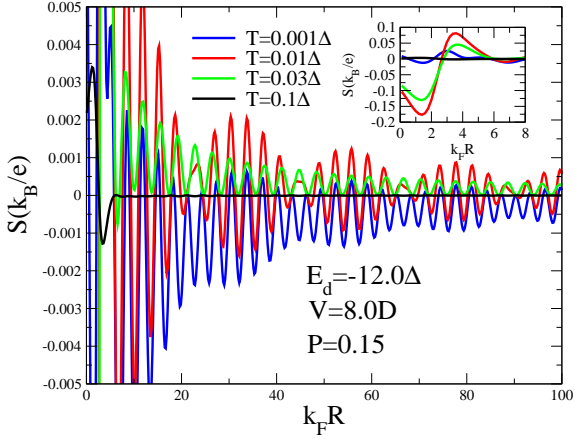} \caption{\label{fig:Pic8}(Color online) Seebeck coefficient (thermopower) $S$ of Eq. (\ref{Seebeck}) in the Kondo regime as function of tip-adatom separation for different values of the temperature. The inset shows the behavior of $S$ for small tip-adatom separations, which allows one to change the sign of the thermopower varying the position of the tip at around the adatom and thus controlling the type of the carriers responsible for the heat flux. The used values of the parameters are: $q_{0}=10$, $E_{d}=-12\Delta$, $V=8.0\Delta$, $P=0.15$.}\end{figure}

\begin{figure}[!]
\includegraphics[width=0.4\textwidth,angle=-90.]{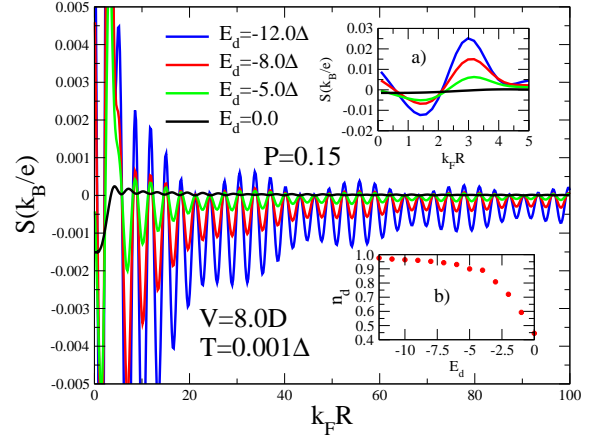} \caption{\label{fig:Pic9}(Color online) Seebeck coefficient (thermopower) $S$ of Eq. (\ref{Seebeck}) in the Kondo regime as function of tip-adatom separation for different values of energy of the adatom level $E_{d},$ which can be tuned by employing an AFM tip capacitively coupled to the adatom as proposed in Ref. {[}\onlinecite{Seridonio1}{]}. Varying the value of $E_d$ one observes a crossover from the intermediate valence regime towards the Kondo limit, which results in the onset of a pronounced oscillation of Seebeck coefficient in the region of small tip-adatom separations shown at the inset (a). Inset (b) shows that the adatom occupation number $n_{d}=n_{d\uparrow}+n_{d\downarrow}$ determined by Eq. (\ref{G00}) attains the unitary limit as a hallmark that the system is within the Kondo regime for $E_{d}=-12.0\Delta$. Parameters used are: $q_{0}=10$, $V=8.0\Delta$, $T=0.001\Delta$, $P=0.15$.}
\end{figure}

\begin{figure}[h]
\includegraphics[width=0.4\textwidth,angle=-90.]{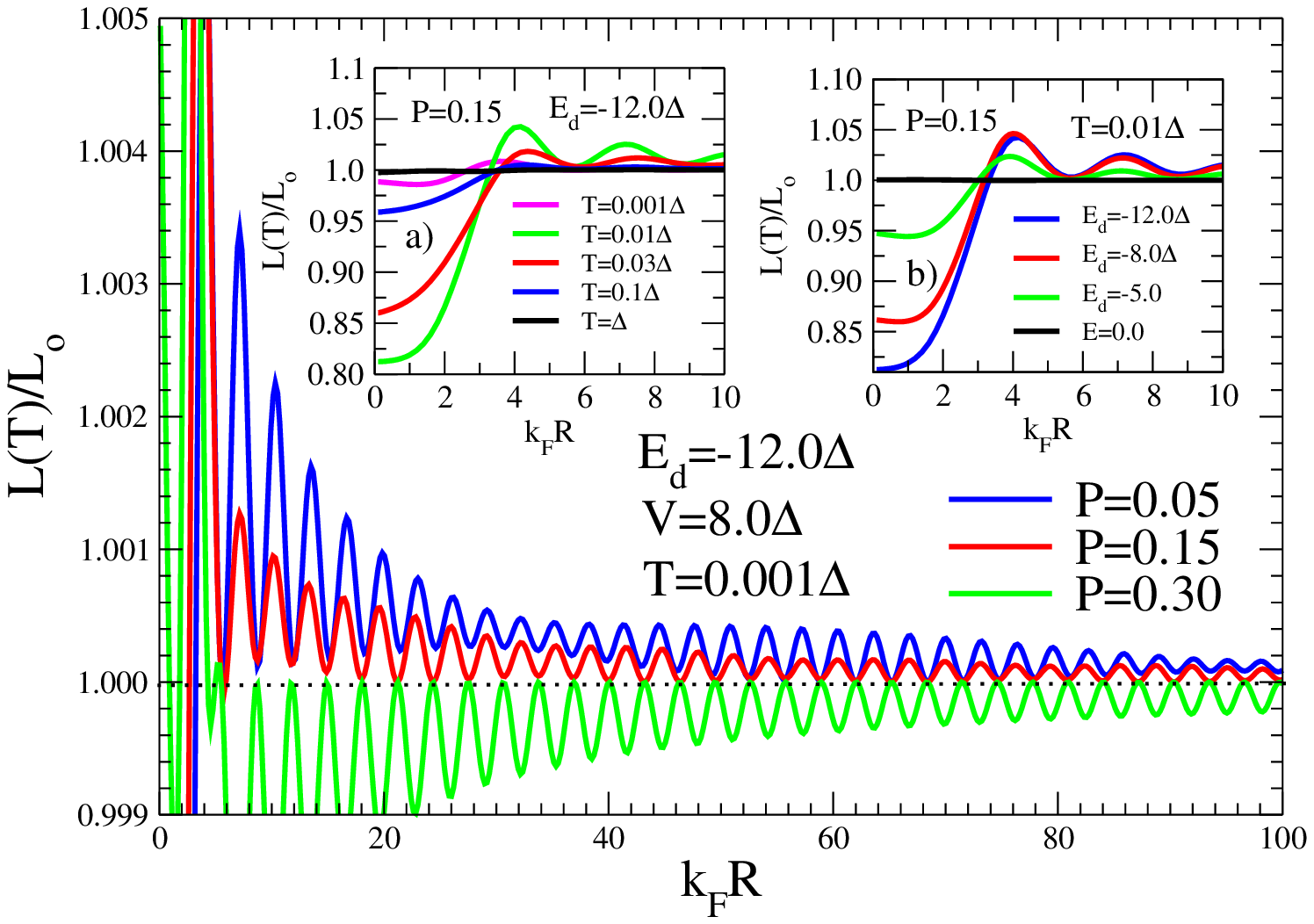} \caption{\label{fig:Pic10}(Color online) Lorenz ratio in the Kondo regime as function of tip-adatom separation for different values of spin polarization. The main plot clearly demonstrates the deviation of the Lorenz number from its standard value $L_{o}=(\frac{\pi^{2}}{3})(\frac{k_{B}}{e})^{2}$ and reveals clear beating pattern. Used parameters are: $q_{0}=10$, $V=8.0\Delta$, $T=0.001\Delta$, $E_{d}=-12.0\Delta$, $P=0.15$. The inset (a) shows $L$ for small tip-adatom separations for different values of the temperature for which deviation from Wiedemann-Franz law is most clearly seen.  The inset (b) shows $L$ for small tip-adatom separations for different values of the adatom energy $E_d$ revealing the crossover from the extreme Kondo limit at $E_{d}=-12.0\Delta$ towards the intermediate valence regime characterized at $E_{d}=0$.}
\end{figure}

Fig. \ref{fig:Pic7} shows the thermopower (Seebeck coefficient) $S$ of Eq. (\ref{Seebeck}) as a function of $k_{F}R$ for different values of spin-polarization $P$ for the host. The sign of the thermopower allows one to determine the type of the carriers responsible for the heat conductance: for $S<0$ they are electrons while for $S>0$ they are holes. In the inset of this figure where the Kondo effect becomes dominant the sign of $S$ changes. As a result, one can tune the type of carrier responsible for the heat conductance just by displacing laterally the STM tip from the adatom site. Therefore, in the Kondo regime the STM device can be used as filter for carriers of the heat flux.

In Fig. \ref{fig:Pic14} we plot the spin Seebeck coefficient (thermopower) $S_{S}$ of Eq. (\ref{eq:Seebeck22}) in the Kondo regime as function of the tip-adatom separation for several polarization values.  According to Figs. \ref{fig:Pic7} and \ref{fig:Pic14}, the magnitude of $S$ and $S_{S}$ are similar. For each polarization value, $S_{S}$ presents near the adatom, pronounced oscillations characterized by amplitudes that increase just by changing the polarization from $P=0.05$ to $P=0.80.$

In the case of spin-accumulation the splitting of the thermopower into spin-up and down components is allowed contrasting to the case of absence of this phenomenon in which such a feature is not possible as Eqs. (\ref{Seebeck}) and (\ref{eq:Seebeck2}) ensure. Remarkably, despite the knowledge of the kind of carriers for the heat flux (holes or electrons) given by the sign of the regular thermopower $S$, the spin thermopower $S_{S}$ provides in addition, the access to the spin degree of freedom (up or down) of these carriers. Thus, the definition encoded by Eq. (\ref{eq:Seebeck22}), which is proportional to the difference $S_{\uparrow}-S_{\downarrow},$ opens the possibility for the knowledge of the heat flux carriers per spin channel $S_{\sigma}.$ As a result, the spin thermopower $S_{S}$ contains simultaneously information about the charge as well as the spin of the aforementioned carriers, thus revealing that the system also behaves as a thermal spin-filter. In Fig. \ref{fig:FFig2} we find within the Kondo regime for $P=0.3$ and $T=0.05\Delta$ with an STM tip faraway the adatom, a situation in which $S_{\uparrow}$ is completely suppressed yielding a heat flux ruled by spin-down electrons as $S_{\downarrow}<0$ and $S_{S}>0.$ In the inset of the same figure we present the profiles of such quantities nearby the adatom where $S_{\uparrow}$ exhibits finite values and competes with  $S_{\downarrow}$.

In Fig. \ref{fig:Pic8} we show the dependence of $S$ in Eq. (\ref{Seebeck}) for different values of the temperature. For very low temperatures ($T=0.001\Delta$), the Seebeck coefficient demonstrates a beating pattern in the range of large tip-adatom separations in full analogy to the electrical and thermal conductances, but as the temperature increases such beats gradually disappear. We point out that in the region of small tip-adatom separations, the thermopower profile is governed by the Kondo effect near the adatom as indicated in the inset of Fig. \ref{fig:Pic8}.

In Fig. \ref{fig:Pic9} we present the thermopower $S$ of Eq. (\ref{Seebeck}) as function of $k_{F}R$ for different  values of the adatom level $E_d$ and fixed spin-polarization $P$. By tuning $E_d$ from the intermediate valence regime, characterized by $E_{d}=0.0$ and $E_{d}=-\Delta$, towards the Kondo regime ($E_{d}=-5.0\Delta$, $E_{d}=-8.0\Delta$ and $E_{d}=-12.0\Delta$), we demonstrate  that the beating pattern is associated to the rising of the Kondo effect. The inset (a) shows the behavior of $S$ in the region of small tip-adatom separations, where the oscillatory pattern arising from the Kondo effect is observed for corresponding values of $E_{d}$. The inset (b) shows that the adatom occupation number $n_{d}=n_{d\uparrow}+n_{d\downarrow}$, determined by Eq. (\ref{G00}), approaches the unitary limit, thus confirming that the system is within the Kondo limit when $E_{d}=-12.0\Delta$.

In Fig. \ref{fig:Pic10} we plot the value of Lorenz number entering into Wiedemann-Franz (WF) law (Eq. (\ref{LL})) in the Kondo regime in units of the Lorenz number for normal metals $L_{o}$ as function of $k_{F}R$. Similar to other thermoelectric coefficients, $L$ reveals characteristic beating pattern. At large distances between the tip and the adatom the amplitude of the beating approaches unity. The inset (a) shows the behavior of $L$ for small tip-adatom separations. One clearly sees that $L/L_0\neq1$ and thus the WF law is violated.  The fulfillment of the WF law is recovered again when the temperature is increased. The violation of the WF law becomes most pronounced in the Kondo regime, when $E_{d}=-12.0\Delta$ and $T=0.01\Delta$, as it is shown in the inset (b). This is a striking result and has been discussed recently in the literature \cite{Kubala,Trocha,Pedro}. The amplitude of the thermopower oscillation near the adatom presents a close relationship with the maximum violation of the WF law: the amplitude of $S$ is maximum, as indicated in the inset of Fig. \ref{fig:Pic8}, just at the temperature where the violation of the WF law is maximum as we can observe in the inset (b) of Fig. \ref{fig:Pic10}.

\section{Conclusions}

\label{sec4}

We have analyzed the beating patterns revealed by thermoelectric coefficients of the STM system and magnetic adatom on conducting surface.  The beating patterns emerge at temperatures close to the Kondo temperature in the range of large tip-adatom separations. In this range, the beats are ruled exclusively by the spin-polarization degree of the ferromagnetic host. For small tip-adatom separations there is an extra dependence on the Fano parameter. Additionally, in this range we have demonstrated the violation of the Wiedemann-Franz law and sign-alternating behavior of the Seebeck coefficient $S$ through charge and spin in the Kondo regime.

The possibility to tune the sign of $S$ opens a way to control the type of the carriers responsible for the heat transfer, as cases $S<0$ and $S>0$ correspond to electrons and holes, respectively. Thus one way to investigate our theoretical predictions is employing the technique of the scanning tunneling microscopy break junction (STMBJ).

\begin{acknowledgments}
This work was supported by the agencies CNPq, PROPe/UNESP, FP7 IRSES projects SPINMET and QOCaN. A. C. Seridonio thanks the University of Iceland and the Nanyang Technological University at Singapore for hospitality.

\end{acknowledgments}

\end{document}